\documentclass[aps,prb,preprint,showpacs]{revtex4-1}
\usepackage{amsfonts}
\usepackage{amsmath}
\usepackage{graphicx}
\usepackage{bm}
\usepackage{amssymb}
\usepackage{dcolumn}
\usepackage{color}

\begin{document}

\title{Diluted magnetic semiconductors with narrow band gaps}

\author{Bo Gu $^{1}$}

\email[Corresponding author: ]{gu.bo@jaea.go.jp}

\author{Sadamichi Maekawa$^{1,2}$}    
 
\affiliation{$^1$ Advanced Science Research Center, Japan Atomic Energy Agency, Tokai 319-1195, Japan \\          
$^2$ ERATO, Japan Science and Technology Agency, Sendai 980-8577, Japan}                            

\date{\today}

\begin{abstract}
We propose a method to realize diluted magnetic semiconductors (DMSs) with p- and n-type carriers by 
choosing host semiconductors with a narrow band gap. 
By employing a combination of the density function theory and quantum Monte Carlo simulation, 
we demonstrate such semiconductors using Mn-doped BaZn$_2$As$_2$, which has a band gap of 0.2 eV. 
In addition, we found a nontoxic DMS Mn-doped BaZn$_2$Sb$_2$, of which the Curie 
temperature T$_c$ is predicted to be higher than that of Mn-doped BaZn$_2$As$_2$, the T$_c$ of which 
was up to 230 K in a recent experiment. 
\end{abstract}

 \maketitle

\section{Introduction}
After the discovery of ferromagnetism in (Ga,Mn)As, diluted magnetic semiconductors (DMSs) have received 
considerable attention owing to potential applications based on the use of both their charge and spin 
degrees of freedom in electronic devices \cite{Ohno,Dietl}. Thus far, the highest Curie temperature of (Ga,Mn)As has 
been T$_c$ = 190 K \cite{Wang}. The substitution of divalent Mn atoms into trivalent Ga sites introduces hole carriers; 
thus, (Ga,Mn)As is a p-type DMS. The valence mismatch between Mn and Ga leads to severely limited chemical solubility 
for Mn in GaAs. Moreover, owing to simultaneous doping of charge and spin induced by Mn substitution, it is difficult 
to individually optimize charge and spin densities. 

To overcome these difficulties, a new type of DMS, i.e., Li(Zn,Mn)As was proposed \cite{Masek} and later 
fabricated with T$_c$ = 50 K \cite{Deng-LiZnAs}. It is based on LiZnAs, a I$-$II$-$V semiconductor. 
Spin is introduced by isovalent (Zn$^{2+}$, Mn$^{2+}$) substitution, which is decoupled from carrier 
doping with excess/deficient Li concentration. Although Li(Zn,Mn)As was proposed as a promising n-type 
DMS with excess Li$^+$, p-type carriers were obtained in the experiment with excess Li. 
The introduction of holes was presumably because of the excess Li$^+$ in substitutional Zn$^{2+}$ sites \cite{Deng-LiZnAs}. 
Later, another I$-$II$-$V DMS, i.e., Li(Zn,Mn)P was reported in an experiment with T$_c$ = 34 K \cite{Deng-LiZnP}. 
Li(Zn,Mn)P with excess Li was determined to be of the p-type as well in the experiment. 
According to first-principles calculations, the reason for this is the same as that for Li(Zn,Mn)As \cite{Deng-LiZnP}. 
Although such p-type I$-$II$-$V DMSs have a few distinct advantages over (Ga,Mn)As, the achievable T$_c$ is much 
lower than that of (Ga,Mn)As. 

Another type of DMS (Ba,K)(Zn,Mn)$_2$As$_2$ was observed in experiments with T$_c$ up to 230 K \cite{Zhao-180,Zhao-230}, 
which is higher than that for (Ga,Mn)As. Based on the semiconductor BaZn$_2$As$_2$, holes were doped by (Ba$^{2+}$, K$^+$) substitutions, and spins by isovalent (Zn$^{2+}$, Mn$^{2+}$) substitutions. It was a p-type DMS. 
Motivated by the high T$_c$, density functional theory (DFT) calculations \cite{Glasbrenner} and photoemission 
spectroscopy experiments \cite{Suzuki-1,Suzuki-2} were conducted to understand the microscopic mechanism of 
ferromagnetism of p-type DMS (Ba,K)(Zn,Mn)$_2$As$_2$.
By contrast, an n-type DMS, i.e., Ba(Zn,Mn,Co)$_2$As$_2$ was recently reported in an experiment 
with T$_c$ $\sim$ 80 K \cite{Man}. In this material, electrons are doped because of the substitution of Zn with Co, 
and spins are generated mainly because of (Zn$^{2+}$, Mn$^{2+}$) substitutions. 

\begin{figure}[tbp]
\includegraphics[width = 8.5 cm]{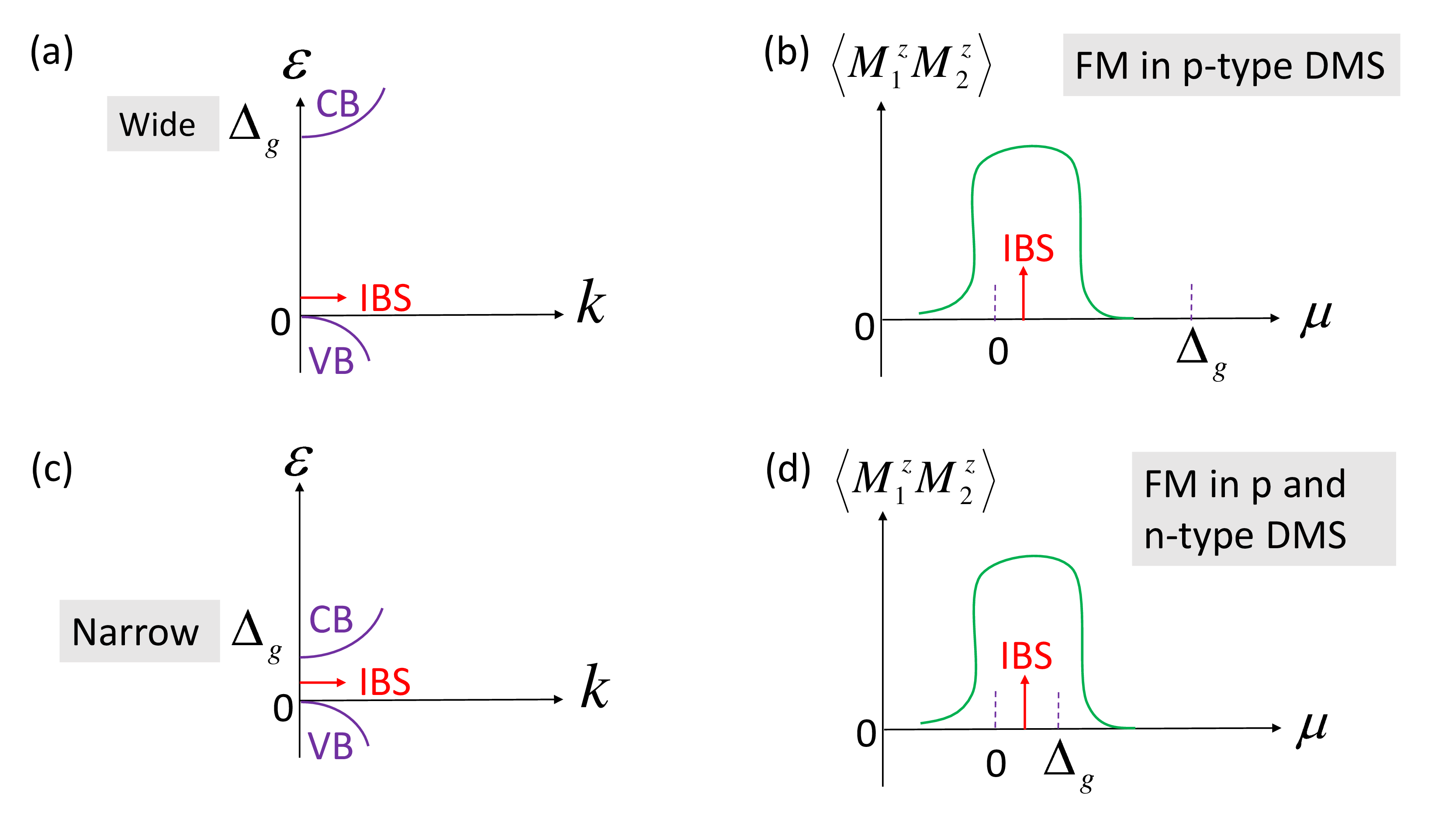}
\caption{ Schematic pictures of an impurity bound state (IBS) and ferromagnetic (FM) coupling in 
diluted magnetic semiconductors (DMSs). 
(a) Host bands $\epsilon (k)$ with a wide band gap $\Delta_{\text{g}}$ between the valence band (VB) and the 
conduction band (CB). The position of the IBS $\omega_{\text{IBS}}$ (arrow) is close to the top of the VB owing 
to strong mixing between the impurity and the VB, and usually no IBS appears below the bottom of the CB because 
of weak mixing between the impurity and the CB \cite{Gu-ZnO,Ohe-GaAs,Gu-MgO}. 
We have 0 $\lesssim$ $\omega_{\text{IBS}}$ $\ll$ $\Delta_{\text{g}}$. 
(b) Magnetic correlation $\langle M_1^zM_2^z\rangle$ between two impurities as a function of the 
chemical potential $\mu$ for case (a). Positive $\langle M_1^zM_2^z\rangle$ denotes FM coupling, 
which can be developed when $\mu$ $\sim$ $\omega_{\text{IBS}}$ \cite{Ichimura,Bulut,Tomoda}. 
Hence, for p-type carriers ($\mu$ $\sim$ 0), FM coupling can be obtained as $\mu$ $\sim$ $\omega_{\text{IBS}}$, 
and for n-type carriers ($\mu$ $\sim$ $\Delta_{\text{g}}$), no magnetic coupling is obtained between 
impurities because $\mu$ $\gg$ $\omega_{\text{IBS}}$ \cite{Gu-ZnO,Ohe-GaAs,Gu-MgO}.  
(c) Similar to case (a), except for a narrow $\Delta_{\text{g}}$. 
By choosing suitable host semiconductors and impurities, 
the condition 0 $\lesssim$ $\omega_{\text{IBS}}$ $\lesssim$ $\Delta_{\text{g}}$ is obtained.
(d) Similar to case (b), except for a narrow $\Delta_{\text{g}}$. 
FM coupling can be achieved for both p-type and n-type carriers 
when $\mu$ $\sim$ $\omega_{\text{IBS}}$. In this study, we describe cases (c) and (d). }
\label{F-schematic}
\end{figure}

In Mn-doped BaZn$_2$As$_2$, why is the ferromagnetic (FM) coupling observed in both p- and n-type cases? 
Why is T$_c$ much lower in the n-type case than that in the p-type case? 
In general, can p- and n-type DMSs be realized? The answers will be helpful for fabricating spin p-n junctions in the future. 
In this study, we attempt to address such issues. In previous studies on DMS materials with wide 
band gap $\Delta_{\text{g}}$, we found that the position of the 
impurity bound state (IBS) $\omega_{\text{IBS}}$ was close to the top of 
the valence band (VB) owing to the strong mixing between the impurity and the VB, 
and usually no IBS appeared below the bottom of the conduction band (CB)
because of weak mixing between the impurity and the CB \cite{Gu-ZnO,Ohe-GaAs,Gu-MgO}. 
Thus, we have 0 $\lesssim$ $\omega_{\text{IBS}}$ $\ll$ $\Delta_{\text{g}}$, 
as shown in Fig. \ref{F-schematic}(a). 
The magnetic correlation $\langle M_1^zM_2^z\rangle$ between two impurities
with FM coupling (positive $\langle M_1^zM_2^z\rangle$) can be determined when the 
chemical potential $\mu$ is tuned to be close to the IBS: $\mu$ $\sim$ $\omega_{\text{IBS}}$ \cite{Ichimura,Bulut,Tomoda}. 
Therefore, for p-type carriers ($\mu$ $\sim$ 0), FM coupling can be obtained as $\mu$ $\sim$ $\omega_{\text{IBS}}$, 
and for n-type carriers ($\mu$ $\sim$ $\Delta_{\text{g}}$), 
no magnetic coupling is obtained between impurities because $\mu$ $\gg$ $\omega_{\text{IBS}}$. 
A schematic diagram describing p-type DMS materials with a wide band gap, 
including (Zn,Mn)O \cite{Gu-ZnO}, (Ga,Mn)As \cite{Ohe-GaAs}, and Mg(O,N) \cite{Gu-MgO},
is shown in Fig. \ref{F-schematic} (b).

Here, we propose a method for realizing p- and n-type DMS. 
The key is choosing host semiconductors with a narrow band gap $\Delta_{\text {g}}$. 
By selecting suitable host semiconductors and impurities, 
the condition 0 $\lesssim$ $\omega_{\text{IBS}}$ $\lesssim$ $\Delta_{\text{g}}$ is satisfied, 
as shown in Fig. \ref{F-schematic}(c). 
We show that for both the p-type ($\mu\sim 0$) and the n-type ($\mu \sim \Delta_{\text{g}}$) cases, 
the condition for developing FM coupling,
that is $\mu$ $\sim$ $\omega_{\text{IBS}}$, can be fulfilled, as shown in Fig. \ref{F-schematic}(d).
 
\section{DFT+QMC Method}
In the following, we realistically calculate the electronic and magnetic properties of the Mn-doped BaZn$_2$As$_2$ DMS, 
which has a narrow band gap $\Delta_{\text{g}}$ (= 0.2 eV) \cite{Zhao-180}. 
We use a combination of the DFT \cite {DFT-1, DFT-2} 
and the Hirsch$-$Fye quantum Monte Carlo (QMC) simulation \cite{QMC}.
Our combined DFT+QMC method can be used for an in-depth treatment of the band structures of materials 
and strong electron correlations of magnetic impurities on an equal footing; 
thus, it can be applied for designing functional semiconductor- \cite {Gu-ZnO,Ohe-GaAs,Gu-MgO} 
and metal-based \cite{Gu-AuFe,Gu-AuPt,Xu-CuIr} materials.
The method involves two calculations steps. 
First, the Haldane$-$Anderson impurity model \cite{Haldane} 
is formulated within the local density approximation for determining the host band
structure and impurity-host mixing. Second, magnetic
correlations of the Haldane-Anderson impurity model at finite
temperatures are calculated using the Hirsch$-$Fye QMC technique \cite{QMC}. 
 
The Haldane$-$Anderson impurity model is defined as follows:
\begin{eqnarray}
  H &=&
  \sum_{\textbf{k},\alpha,\sigma}[\epsilon_{\alpha}(\textbf{k})-\mu]
  c^{\dag}_{\textbf{k}\alpha\sigma}c_{\textbf{k}\alpha\sigma} 
  +\sum_{\textbf{k},\alpha,\textbf{i},\xi,\sigma}(V_{\textbf{i}\xi\textbf{k}\alpha }
 d^{\dag}_{\textbf{i}\xi\sigma} c_{\textbf{k}\alpha\sigma} \notag\\
   &+& h.c.) + (\epsilon_d-\mu)\sum_{\textbf{i},\xi,\sigma}
   d^{\dag}_{\textbf{i}\xi\sigma}d_{\textbf{i}\xi\sigma}
   + U\sum_{\textbf{i},\xi}n_{\textbf{i}\xi\uparrow}n_{\textbf{i}\xi\downarrow},
   \label{E-Ham}
\end{eqnarray}
where $c^{\dag}_{\textbf{k}\alpha\sigma}$
($c_{\textbf{k}\alpha\sigma}$) is the creation (annihilation)
operator for a host electron with wave vector $\textbf{k}$ and spin
$\sigma$ in the VB ($\alpha = v$) or the CB ($\alpha = c$), and $d^{\dag}_{\textbf{i}\xi\sigma}$
($d_{\textbf{i}\xi\sigma}$) is the creation (annihilation) operator
for a localized electron at impurity site $\textbf{i}$ in orbital
$\xi$ and spin $\sigma$ with
$n_{\textbf{i}\xi\sigma}=d^{\dag}_{\textbf{i}\xi\sigma}d_{\textbf{i}\xi\sigma}$.
Here, $\epsilon_{\alpha}(\textbf{k})$ is the host band dispersion,
$\mu$ is the chemical potential, $V_{\textbf{i}\xi\textbf{k}\alpha}$
denotes mixing between the impurity and the host, $\epsilon_d$ is the
impurity $3d$ orbital energy, and $U$ is the on-site Coulomb
repulsion of the impurity. 
Considering the condition of Hund coupling $J_H\ll U$, 
$J_H$ is neglected and the single-orbital approximation is used to 
describe the magnetic sates of impurities.  

\section{Results for $Ba(Zn,Mn)_2As_2$}
\begin{figure}[tbp]
\includegraphics[width = 8.5 cm]{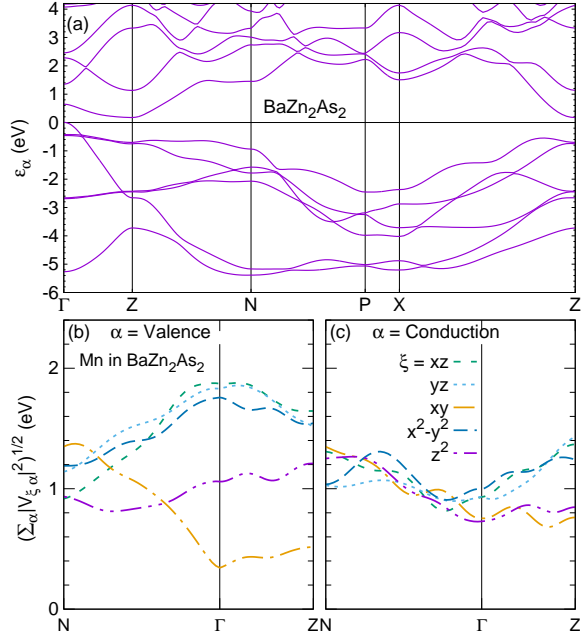}
\caption{Host band and mixing parameters of Mn-doped BaZn$_2$As$_2$. 
(a) Energy bands $\epsilon_{\alpha}$ of host BaZn$_2$As$_2$,
which has space group I4/mmm. 
An indirect band gap of 0.2 eV was obtained by DFT calculations, which agrees well with the experimental value \cite{Zhao-180}. 
The mixing function between the $\xi$ orbitals of an Mn impurity and BaZn$_2$As$_2$ hosts (b) valence bands
and (c) conduction bands. }
\label{F-band-mix}
\end{figure}

The parameters $\epsilon_{\alpha}(\textbf{k})$ and $V_{\textbf{i}\xi\textbf{k}\alpha}$
are obtained by DFT calculations using the Wien2k package \cite{Wien2k}.
To reproduce the experimental narrow band gap of 0.2 eV in BaZn$_2$As$_2$ \cite{Zhao-180}, 
we use the modified Becke$-$Johnsom exchange potential (mBJ) \cite{mBJ}, 
which has been implemented in the Wien2k package. 
The obtained energy band $\epsilon_{\alpha}$ (\textbf{k}) is shown in Fig. \ref{F-band-mix} (a), 
where BaZn$_2$As$_2$ has space group I4/mmm. We obtained an indirect gap band $\Delta_{\text{g}}$ = 0.2 eV, 
which is in good agreement with the experimental \cite {Zhao-180} and previous calculated \cite{Suzuki-1, Shein} values.

The mixing parameter between the $\xi$ orbitals of an Mn impurity and the BaZn$_2$As$_2$ host is 
defined as $V_{\textbf{i}\xi\textbf{k}\alpha}$$\equiv$$\langle\varphi_{\xi}
(\textbf{i})|H|\Psi_{\alpha}(\textbf{k})\rangle$$\equiv$$\frac{1}{\sqrt{N}}e^{i \textbf{k}\cdot
\textbf{i}}V_{\xi\alpha }(\textbf{k})$, which can be expressed as 
\begin{eqnarray}
V_{\xi\alpha }(\textbf{k}) = \sum_{o,\textbf{n}}e^{i \textbf{k}\cdot
(\textbf{n}-\textbf{i})}a_{\alpha o}(\textbf{k})
\langle\varphi_{\xi}(\textbf{i})|H|\varphi_{o}(\textbf{n})\rangle,
\label{E-Mix}
\end{eqnarray}
where $\varphi_{\xi} (\textbf{i})$ is the impurity $3d$ state at site $\textbf{i}$, 
and $\Psi_{\alpha}(\textbf{k})$ is the host state with wave vector $\textbf{k}$ and band index $\alpha$,
which is expanded by atomic orbitals $\varphi_{o}(\textbf{n})$ having
orbital index $o$ and site index $\textbf{n}$. Here, $N$ is the
total number of host lattice sites, and $a_{\alpha o}(\textbf{k})$
is an expansion coefficient. To obtain the mixing integrals of
$\langle\varphi_{\xi}(\textbf{i})|H|\varphi_{o}(\textbf{n})\rangle
$, we consider a supercell Ba$_8$Zn$_{15}$MnAs$_{16}$, which is comprised of 2x2x2 primitive cells,
where each primitive cell consists of a BaZn$_2$As$_2$, and a Zn atom is replaced by an Mn atom.  
The results of the mixing function $V_{\xi\alpha }(\textbf{k})$ are shown in Fig. \ref{F-band-mix} (b) for valence bands, 
and in Fig. \ref{F-band-mix} (c) for conduction bands. 

The parameters $U$ and $\epsilon_d$ are determined as follows. 
For (Ga,Mn)As, the reasonable parameters are estimated as 
$U$ = 4 eV and $\epsilon_d$ = -2 eV \cite{Ohe-GaAs}.
A recent resonance photoemission spectroscopy experiment showed that the Mn $3d$ partial density of states 
in (Ba,K)(Zn,Mn)$_2$As$_2$ and (Ga,Mn)As are quite similar, excepted that the peak of (Ga,Mn)As is approximately 0.4 eV deeper 
than that of (Ba,K)(Zn,Mn)$_2$As$_2$ \cite{Suzuki-1}. 
Thus, the reasonable parameters of Mn-doped BaZn$_2$As$_2$ are $U$ = 4 eV and $\epsilon_d$ = -1.5 eV.  
On the basis of the parameters obtained above, magnetic correlations of the impurities are calculated using the Hirsch$-$Fye QMC
technique with more than 10$^{6}$ Monte Carlo sweeps and a Matsubara time step $\Delta\tau$ = 0.25.

\begin{figure}[tbp]
\includegraphics[width = 8.5 cm]{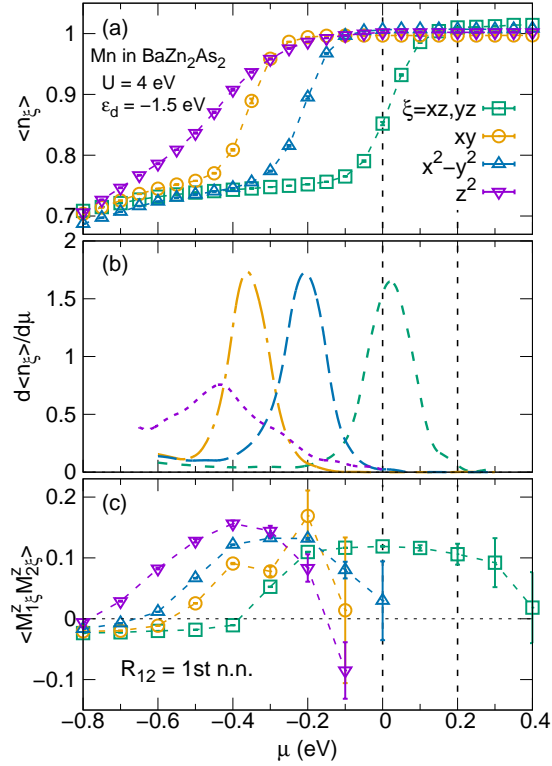}
\caption{For Mn-doped BaZn$_2$As$_2$, chemical potential $\mu$ dependence of (a) occupation number $\langle n_{\xi}\rangle$ of $\xi$ orbital of an Mn impurity, (b) partial density of state $d\langle n_{\xi}\rangle/d\mu$, 
and (c) magnetic correlation $\langle M_{1\xi}^zM_{2\xi}^z\rangle$ between the $\xi$ orbitals of two Mn impurities with fixed distance 
$R_{12}$ of the first-nearest neighbor, where the temperature is 360 K. The top of the VB is 0, and the bottom of the CB is 0.2 eV.}
\label{F-ibs}
\end{figure}

Figure \ref{F-ibs} (a) shows a plot of the occupation number $\langle n_{\xi}\rangle$ of a $\xi$ orbital 
of an Mn impurity in BaZn$_2$As$_2$ against the chemical potential $\mu$ at 360 K. 
The top of the VB was taken to be 0, and the bottom of the CB to be 0.2 eV.
Operator $n_{\xi}$ is defined as follows:
\begin{equation}
n_{\xi} = n_{\textbf{i}\xi\uparrow} + n_{\textbf{i}\xi\downarrow}.
\end{equation}
The orbitals $xz$ and $yz$ of Mn subsitutional impurities at the Zn site degenerate owing to the crystal field of BaZn$_2$As$_2$, 
which has a group space of I4/mmm \cite{Zhao-180}. 
Sharp increases in $n_{\xi}$ are observed around -0.5, -0.4, -0.2, and 0.0 eV for the orbitals $\xi$ = $z^2$, $xy$, $x^2-y^2$, 
and $xz(yz)$, respectively. 
This implies the existence of an IBS at this energy $\omega_{\text{IBS}}$ \cite{Gu-ZnO,Ohe-GaAs,Gu-MgO,Ichimura,Bulut,Tomoda}.
In order to make the IBS clearer, we show the partial density of state of an Mn impurity, 
$d\langle n_{\xi}\rangle/d\mu$, in Fig. \ref{F-ibs} (b).  
The peaks in $d\langle n_{\xi}\rangle/d\mu$ correspond to the positions of IBS. 
Figure \ref{F-ibs} (c) shows the magnetic correlation $\langle M_{1\xi}^zM_{2\xi}^z\rangle$ between the $\xi$ orbitals of two Mn 
impurities with fixed distance $R_{12}$ of the first-nearest neighbor. 
The operator $M^z_{\textbf{i}\xi}$ of the $\xi$ orbital at impurity site $\textbf{i}$ is defined as follows:
\begin{equation}
M^z_{\textbf{i}\xi} = n_{\textbf{i}\xi\uparrow} - n_{\textbf{i}\xi\downarrow}.
\end{equation}
For each $\xi$ orbital, FM coupling is obtained when the chemical potential $\mu$ is close to the IBS position, 
and FM correlations become weaker and eventually disappear when $\mu$ moves away from the IBS. 
This role of the IBS in determining the strength of FM correlations between impurities is consistent with 
the Hartree$-$Fock and QMC results of various DMS systems \cite{Gu-ZnO,Ohe-GaAs,Gu-MgO,Ichimura,Bulut,Tomoda}.

\begin{figure}[tbp]
\includegraphics[width = 8.5 cm]{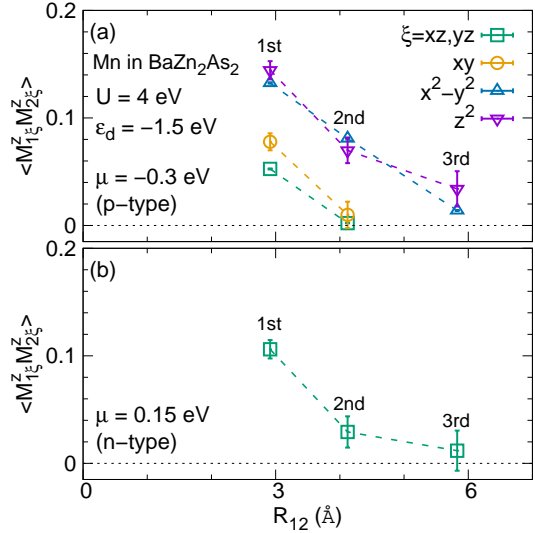}
\caption{For Mn-doped BaZn$_2$As$_2$, the distance $R_{12}$ dependence of magnetic correlation $\langle M_{1\xi}^zM_{2\xi}^z\rangle$ between the $\xi$ orbitals of two Mn impurities for the (a) p-type case with chemical potential $\mu$ = -0.3 eV and (b) n-type case with $\mu$ = 0.15 eV, where temperature is 360 K. The first-, second-, and third-nearest neighbors of $R_{12}$ are noted. }
\label{F-c2}
\end{figure}

For Mn-doped BaZn$_2$As$_2$ with p-type carriers, a recent angle-resolved photoemission spectroscopy (ARPES) experiment showed that 
the Fermi level ($\mu$) is below the top of the VB 
by several tenths of an eV and a non-dispersive Mn 3$d$ impurity band is present slightly below the Fermi level \cite{Suzuki-2}. 
On the basis of the results in Fig. \ref{F-ibs} (a), we take $\mu$ = -0.3 eV as an estimate for the p-type case. 
We argue that the IBS of orbitals $xy$ and $z^2$, whose positions are below the $\mu$ = -0.3 eV, can account for the non-dispersive Mn 3$d$ impurity band 
below the Fermi level observed in the ARPES experiment. 
Figure \ref{F-c2} (a) shows the distance $R_{12}$ dependence of the magnetic correlation $\langle M_{1\xi}^zM_{2\xi}^z\rangle$ between the $\xi$ orbitals of two Mn impurities for the p-type case with $\mu$ = -0.3 eV. Long-range FM coupling up to approximately 6 \AA (the third nearest neighbor) is obtained for the orbitals $\xi$ = $x^2-y^2$ and $z^2$, while short-range FM coupling is obtained for the other three orbitals.  
Thus, our theoretical results are consistent with the FM observed in the experiment involving Mn-doped BaZn$_2$As$_2$ with p-type carriers.  

For Mn-doped BaZn$_2$As$_2$ with n-type carriers, a recent experiment showed FM coupling below T$_c$ = 80 K \cite{Man}. Because no information about the Fermi level has been reported, we take $\mu$ = 0.15 eV as an estimate for the n-type case, which is below the bottom of the CB by 0.05 eV. As shown in Fig. \ref{F-c2} (b), 
long-range FM coupling up to approximately 6 \AA (the 3rd nearest neighbor) is obtained for the orbitals $\xi$ = $xz$ and $yz$. No FM is obtained for the other three orbitals, shown in Fig. \ref{F-ibs} (b) as well.  
A Comparison of Figs. \ref{F-c2}(a) and \ref{F-c2}(b) shows that the magnitude of FM coupling $\langle M_{1\xi}^zM_{2\xi}^z\rangle$ in the n-type case is smaller than that in the p-type case, which can qualitatively explain why the T$_c$ in the n-type case \cite {Man} is lower than that in the 
p-type case \cite {Zhao-180, Zhao-230} in the experiments. 

\begin{figure}[tbp]
\includegraphics[width = 8.5 cm]{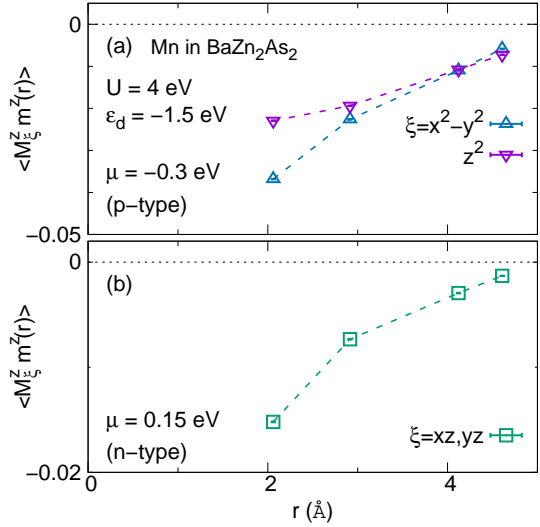}
\caption{For Mn-doped BaZn$_2$As$_2$, the distance $r$ dependence of magnetic correlation $\langle M_{\xi}^zm^z(r)\rangle$ between the $\xi$ orbitals of Mn impurity at site origin and the host electron at site $\bf{r}$ for the (a) p-type case with chemical potential $\mu$ = -0.3 eV and (b) n-type case with $\mu$ = 0.15 eV, where temperature is 360 K.}
\label{F-host}
\end{figure}

To understand the long-range FM correlation function $\langle
M^z_{1}M^z_{2}\rangle$ between two Mn impurities in the BaZn$_2$As$_2$ host, 
we have calculated the impurity-host magnetic correlation function
$\langle M^z_{\xi}m^z(\textbf {r})\rangle$. Here, $\textbf {r}$ is the
site of the host electron and the impurity Mn is located at site $\textbf
{r}$ = 0. The magnetization $m^z(\textbf {r})$ of the host electron at site $\textbf{r}$ is defined as
\begin{equation}
m^z(\textbf {r})=\sum_{\alpha}( n_{\alpha \textbf {r}\uparrow}
-n_{\alpha \textbf {r}\downarrow}),
\end{equation}
where $n_{\alpha\textbf{r}\sigma}=c^{\dag}_{\alpha\textbf{r}\sigma}
c_{\alpha\textbf{r}\sigma}$ is the number operator for host
electrons with band index $\alpha$ and site $\textbf {r}$ and spin
$\sigma$. In Fig. \ref{F-host}(a), for p-type carriers with $\mu$ = -0.3 eV,
the long-range antiferromagnetic (AFM) correlation is obtained between the orbitals 
$\xi$ = $x^2-y^2$ and $z^2$ of Mn impurity and host electrons. 
In Fig. \ref{F-host}(b), for n-type carriers with $\mu$ = 0.15 eV,
the long-range AFM correlation is obtained between the orbitals 
$\xi$ = $xz$ and $yz$ of Mn impurity and host electrons. 
Thus, the long-range FM coupling between impurities is mediated by the polarization of host electron spin. 
Such carrier-mediated FM is already discussed in previous DMS materials with a wide band gap, 
such as (Zn,Mn)O \cite{Gu-ZnO}, (Ga,Mn)As \cite{Ohe-GaAs}, and Mg(O,N) \cite{Gu-MgO}. 

\section{Results for $Ba(Zn,Mn)_2Sb_2$}
\begin{figure}[tbp]
\includegraphics[width = 8.5 cm]{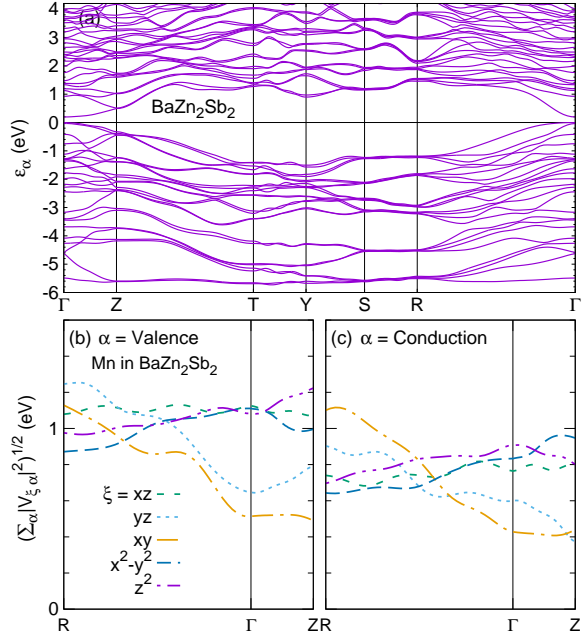}
\caption{Similar to Fig. \ref{F-band-mix}, with the exception that BaZn$_2$As$_2$ is replaced by BaZn$_2$Sb$_2$,
which has a different space group Pnma. 
A direct band gap of 0.2 eV is obtained by DFT calculations, which is in good agreement with the experimental value \cite{Madsen}. }
\label{F-bazn2sb2-mix}
\end{figure}

We made similar calculations for Mn-doped BaZn$_2$Sb$_2$,
where a distinct advantage was the replacement of As with nontoxic Sb.
BaZn$_2$Sb$_2$, too, has a narrow band gap $\Delta_{\text{g}}$ = 0.2 eV, but a different space group Pnma \cite{Madsen}.   
A direct band gap of 0.2 eV was obtained by the DFT calculation as shown in Fig. \ref{F-bazn2sb2-mix} (a), 
which agrees well with the experimental value.

\begin{figure}[tbp]
\includegraphics[width = 8.5 cm]{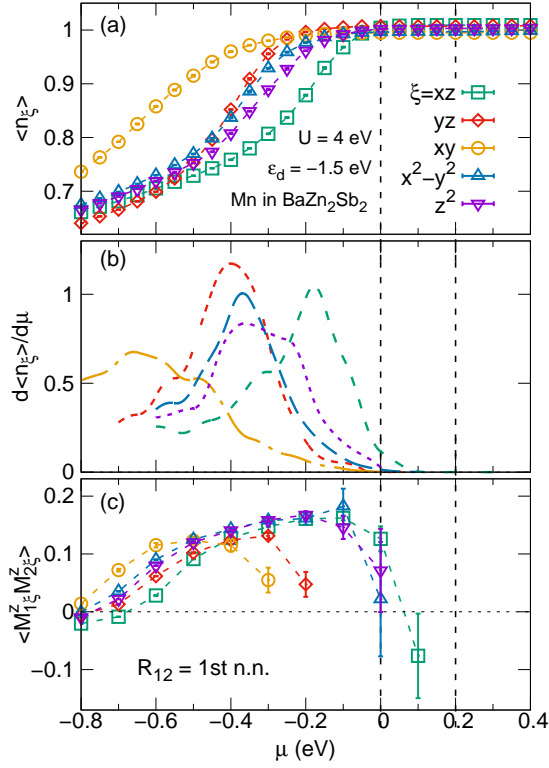}
\caption{Similar to Fig. \ref{F-ibs}, except BaZn$_2$As$_2$ is replaced by BaZn$_2$Sb$_2$.  }
\label{F-ibs-bazn2sb2}
\end{figure}

Figure \ref{F-ibs-bazn2sb2} (a) shows the occupation number $\langle n_{\xi}\rangle$ of the $\xi$ orbital 
of the Mn impurity in BaZn$_2$Sb$_2$ versus chemical potential $\mu$ at temperature 360 K. 
The $3d$ orbitals of Mn did not degenerate owing to the low symmetry of the crystal field of BaZn$_2$Sb$_2$. 
Sharp increases in $\langle n_{\xi}\rangle$, which imply the position of IBS $\omega_{\text{IBS}}$, were observed around -0.6 eV for the $xy$ orbital, 
-0.4 eV for the $yz$, $x^2-y^2$, and $z^2$ orbitals, and -0.2 eV for the $xz$ orbital. 
The IBS can be seen more clearly in the partial density of state of an Mn impurity, 
$d\langle n_{\xi}\rangle/d\mu$, in Fig. \ref{F-ibs-bazn2sb2} (b). 
The peaks in $d\langle n_{\xi}\rangle/d\mu$ correspond to the positions of the IBS.  
Figure \ref{F-ibs-bazn2sb2} (c) shows the magnetic correlation $\langle M_{1\xi}^zM_{2\xi}^z\rangle$ between the $\xi$ orbitals of two Mn impurities with fixed distance $R_{12}$ as the first nearest neighbor. 
The role of the IBS in determining the strength of the FM correlations between impurities is the same as that discussed for Mn-doped BaZn$_2$As$_2$ in Fig. \ref{F-ibs}.  

\begin{figure}[tbp]
\includegraphics[width = 8.5 cm]{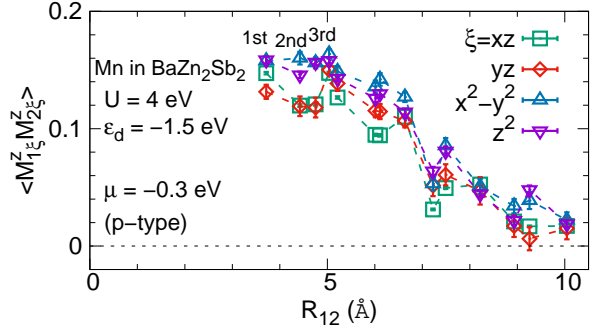}
\caption{Similar to Fig. \ref{F-c2}(a), except BaZn$_2$As$_2$ is replaced with BaZn$_2$Sb$_2$. }
\label{F-c2-bazn2sb2}
\end{figure}

For Mn-doped BaZn$_2$Sb$_2$ with p-type carriers, we take $\mu$ = -0.3 eV, the same value as that used for Mn-doped BaZn$_2$As$_2$ with p-type carriers. 
Figure \ref{F-c2-bazn2sb2} shows the distance $R_{12}$ dependence of the magnetic correlation $\langle M_{1\xi}^zM_{2\xi}^z\rangle$ between the $\xi$ orbitals of two Mn impurities for the p-type case. Long-range FM coupling up to approximately 10 \AA (the 14th nearest neighbor) was obtained for the $\xi$ = $xz$, $yz$, $x^2-y^2$, and $z^2$ orbitals, while relatively short-range FM coupling is obtained for the $xy$ orbital. 
This is considerably longer than 6 \AA (the third nearest neighbor) obtained for Mn doped BaZn$_2$As$_2$ with p-type carriers, as shown in Fig. \ref{F-c2} (a). Such long-range FM coupling arises from the short distance between the neighboring Zn sites in BaZn$_2$Sb$_2$, as is clear from comparison of the first-, second-, and third-nearest neighbors in Fig. \ref{F-c2}(a) and those neighbors in Fig. \ref{F-c2-bazn2sb2}, respectively.  
We predict that the T$_c$ of Mn-doped BaZn$_2$Sb$_2$ with p-type carriers should be higher than that of Mn-doped BaZn$_2$As$_2$ with p-type carriers, in which 
 T$_c$ = 230 K was reported in a recent experiment \cite{Zhao-230}. 

For Mn-doped BaZn$_2$Sb$_2$ with n-type carriers, we take $\mu$ = 0.15 eV, the same value as that used for Mn-doped BaZn$_2$As$_2$ with n-type carriers. No FM coupling is obtained with $\mu$ = 0.15 eV.
This is because $\mu$ = 0.15 eV is far from the IBS position $\omega_{\text{IBS}}\approx$ -0.2 eV of the $xz$ orbital, 
as shown in Figs. \ref{F-ibs-bazn2sb2}(a) and \ref{F-ibs-bazn2sb2}(b).  
It is consistent with previous studies that no magnetic coupling is obtained between impurities when 
$\mu \gg \omega_{\text{IBS}}$\cite{Gu-ZnO,Ohe-GaAs,Gu-MgO}. 

\section{Discussion on uncertainty of model parameters}

In the above QMC calculations, we fix the model parameters 
of impurity level $\epsilon_d$ = -1.5 eV and Coulomb repulsion $U$ = 4 eV, 
which are reasonable values for Mn-doped BaZn$_2$As$_2$ and BaZn$_2$Sb$_2$ as discussed in Sec. III. 
In this section, we will discuss how the uncertainty of these values affects the outcome of the calculations.

\begin{figure}[tbp]
\includegraphics[width = 8.5 cm]{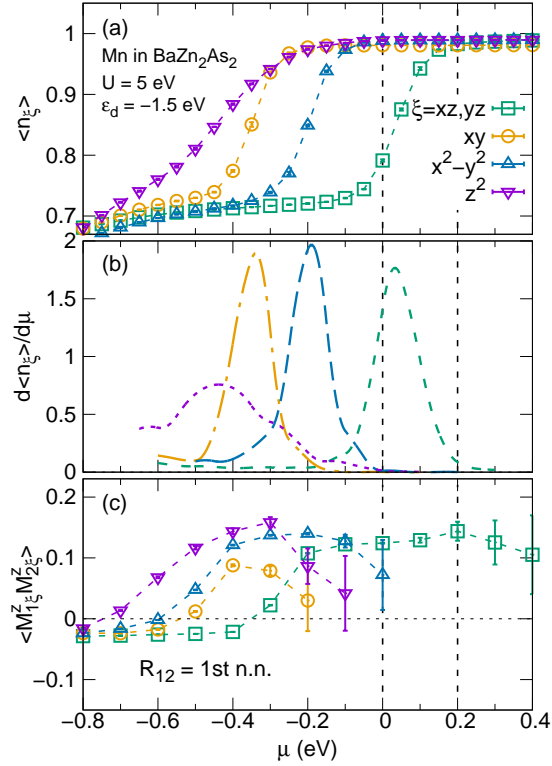}
\caption{Similar to Fig. \ref{F-ibs}, except Coulomb repulsion parameter $U$ = 4 eV is replaced by a larger value $U$ = 5 eV.  }
\label{F-ibs-u5}
\end{figure}
	
For Mn-doped BaZn$_2$As$_2$ with the same impurity level parameter $\epsilon_d$ = -1.5 eV 
and a larger Coulomb repulsion parameter $U$ = 5 eV, 
the occupation number $\langle n_{\xi}\rangle$,
the partial density of state $d\langle n_{\xi}\rangle/d\mu$ of the $\xi$ orbital of an Mn impurity,
and the magnetic correlation $\langle M_{1\xi}^zM_{2\xi}^z\rangle$ between the $\xi$ orbitals of two Mn 
impurities with fixed distance of the first-nearest neighbor are shown in Figs. \ref{F-ibs-u5}(a)-\ref{F-ibs-u5}(c), respectively.  
Compared with the results obtained with parameters $\epsilon_d$ = -1.5 eV and $U$ = 4 eV in Fig. \ref{F-ibs}, 
no essential difference is observed. 
	
\begin{figure}[tbp]
\includegraphics[width = 8.5 cm]{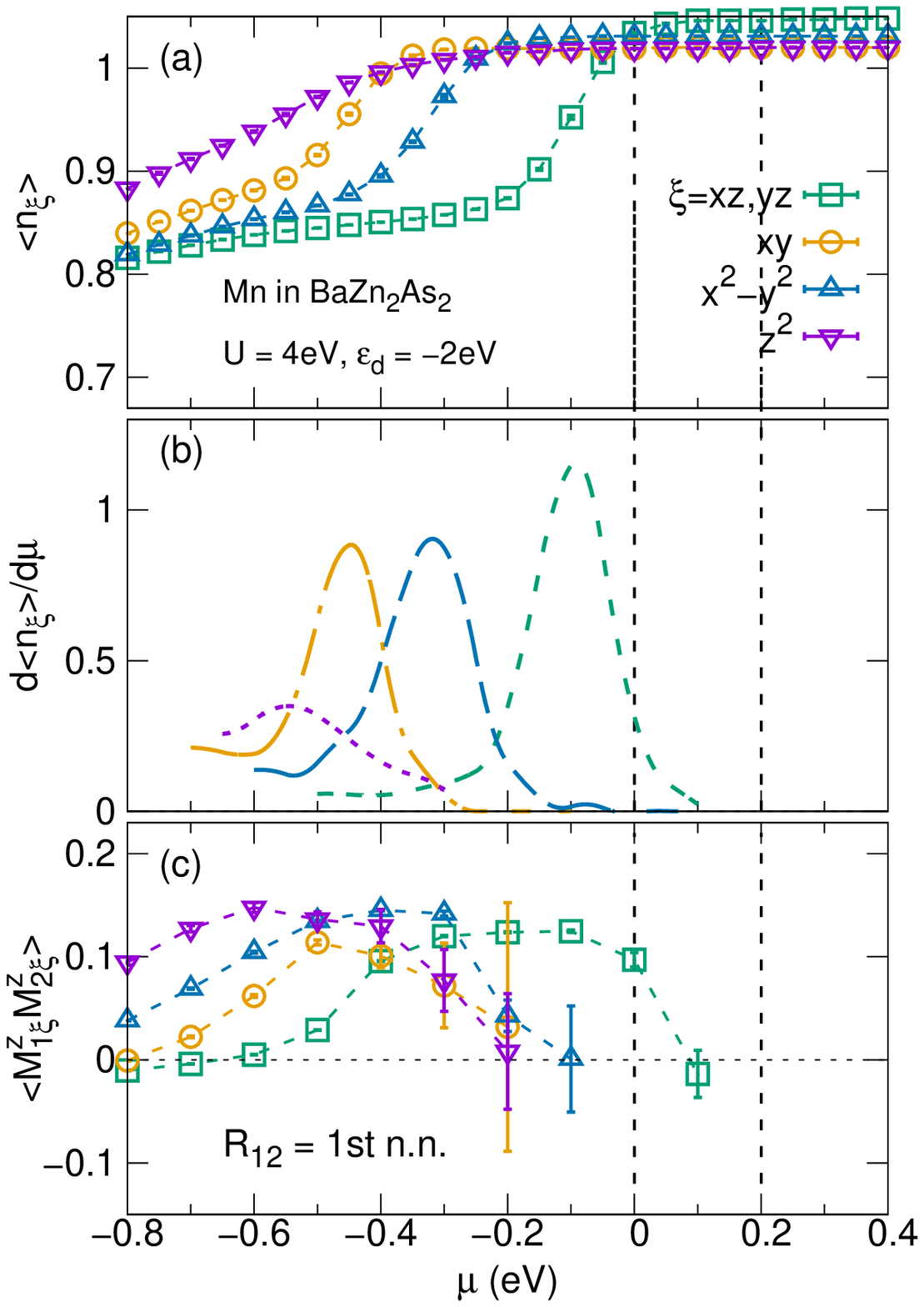}
\caption{Similar to Fig. \ref{F-ibs}, except impurity level parameter $\epsilon_d$ = -1.5 eV is replaced by a deeper value $\epsilon_d$ = -2 eV. }
\label{F-ibs-em2}
\end{figure}

For Mn-doped BaZn$_2$As$_2$ with a deeper impurity level parameter $\epsilon_d$ = -2 eV 
and the same Coulomb repulsion parameter $U$ = 4 eV,
$\langle n_{\xi}\rangle$, $d\langle n_{\xi}\rangle/d\mu$,
and $\langle M_{1\xi}^zM_{2\xi}^z\rangle$ are shown in Figs. \ref{F-ibs-em2}(a)-\ref{F-ibs-em2}(c), respectively.  
Compared with the results in Fig. \ref{F-ibs}, 
the IBS positions $\omega_{\text{IBS}}$ of $\xi$ orbitals of Mn impurity shift down by about 0.1 eV. 
As a result, the FM correlation $\langle M_{1\xi}^zM_{2\xi}^z\rangle$ is obtained for p-type carriers with $\mu$ = -0.3 eV, 
while no FM correlation $\langle M_{1\xi}^zM_{2\xi}^z\rangle$ is obtained for n-type carriers with $\mu$ = 0.15 eV.  
This result does not agree with the recent experiment of Mn-doped BaZn$_2$As$_2$ with n-type carriers, where FM coupling is observed below T$_c$ = 80 K \cite{Man}. Thus, the impurity level parameter $\epsilon_d$ = -2 eV may be too deep for Mn-doped BaZn$_2$As$_2$, as we have also discussed in Sec. III.   

\begin{figure}[tbp]
\includegraphics[width = 8.5 cm]{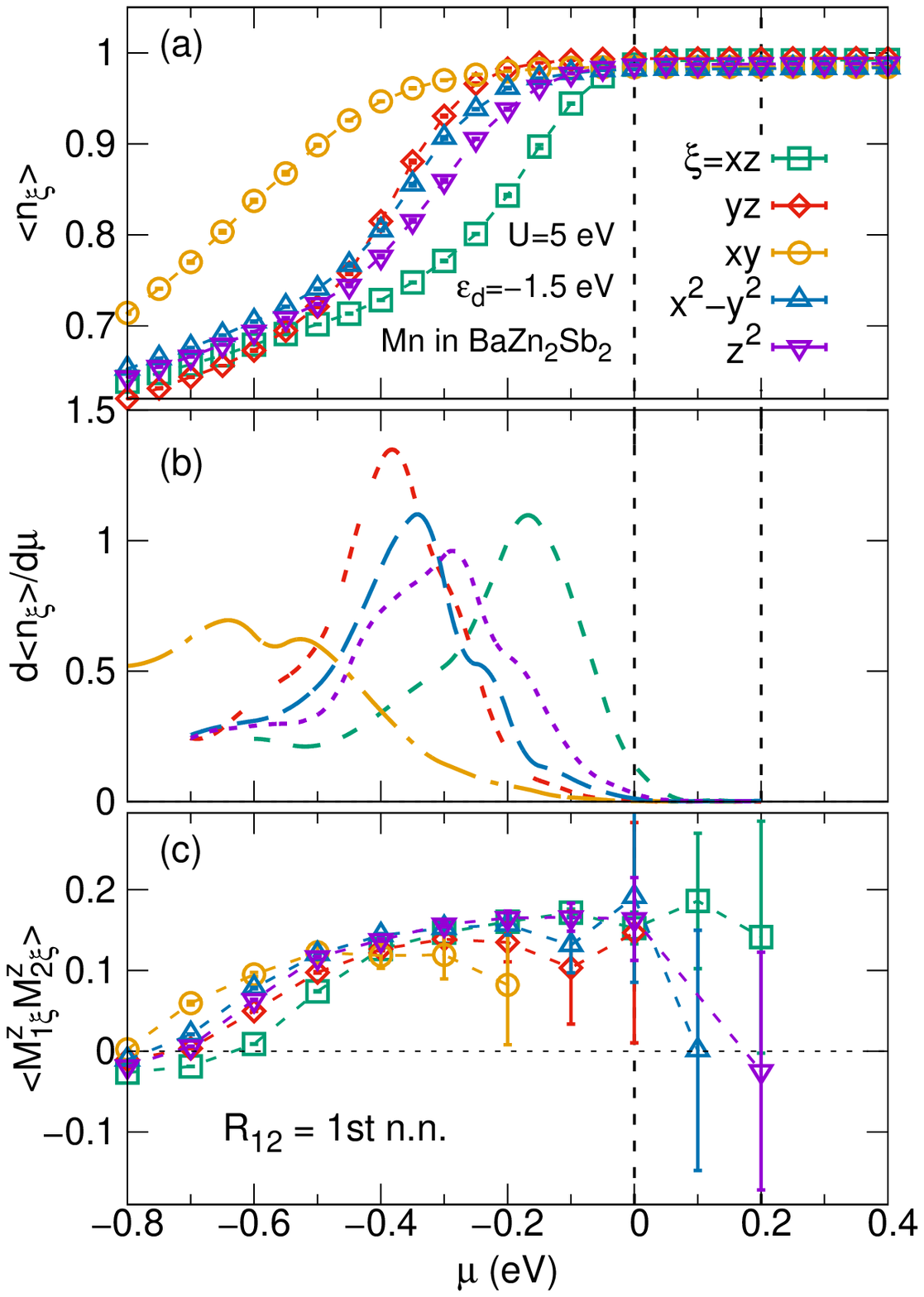}
\caption{Similar to Fig. \ref{F-ibs-bazn2sb2}, except Coulomb repulsion parameter $U$ = 4 eV is replaced by a larger value $U$ = 5 eV. }
\label{F-ibs-bazn2sb2-u5}
\end{figure}

For Mn-doped BaZn$_2$Sb$_2$ with the same impurity level parameter $\epsilon_d$ = -1.5 eV 
and a larger Coulomb repulsion parameter $U$ = 5 eV, 
$\langle n_{\xi}\rangle$, $d\langle n_{\xi}\rangle/d\mu$,
and $\langle M_{1\xi}^zM_{2\xi}^z\rangle$ are shown in Figs. \ref{F-ibs-bazn2sb2-u5}(a)-\ref{F-ibs-bazn2sb2-u5}(c), respectively.  
Compared with the results obtained with parameters $\epsilon_d$ = -1.5 eV and $U$ = 4 eV in Fig. \ref{F-ibs-bazn2sb2}, 
no essential difference is observed. 

\begin{figure}[tbp]
\includegraphics[width = 8.5 cm]{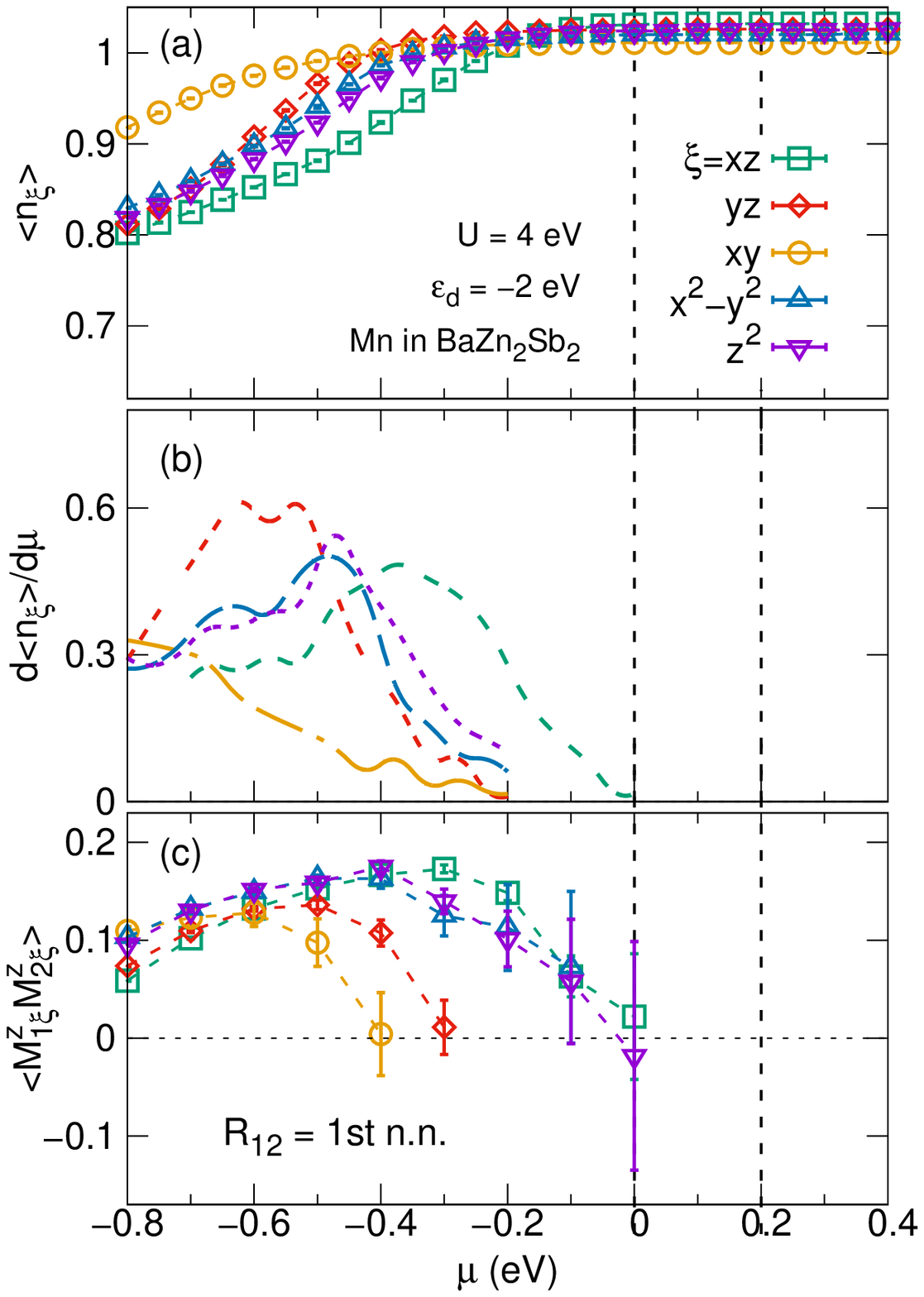}
\caption{Similar to Fig. \ref{F-ibs-bazn2sb2}, except impurity level parameter $\epsilon_d$ = -1.5 eV is replaced by
a deeper value $\epsilon_d$ = -2 eV. }
\label{F-ibs-bazn2sb2-em2}
\end{figure}

For Mn-doped BaZn$_2$Sb$_2$ with a deeper impurity level parameter $\epsilon_d$ = -2.0 eV 
and the same Coulomb repulsion parameter $U$ = 4 eV, 
$\langle n_{\xi}\rangle$, $d\langle n_{\xi}\rangle/d\mu$,
and $\langle M_{1\xi}^zM_{2\xi}^z\rangle$ are shown in Figs. \ref{F-ibs-bazn2sb2-em2}(a)-\ref{F-ibs-bazn2sb2-em2}(c), respectively.  
Compared with the results in Fig. \ref{F-ibs-bazn2sb2}, 
the IBS positions $\omega_{\text{IBS}}$ of $\xi$ orbitals of Mn impurity shift down by about 0.1 eV.
The FM correlation $\langle M_{1\xi}^zM_{2\xi}^z\rangle$ is 
obtained for p-type carriers with $\mu$ = -0.3 eV, and no FM correlation $\langle M_{1\xi}^zM_{2\xi}^z\rangle$ is 
obtained for n-type carriers with $\mu$ = 0.15 eV. The conclusion is unchanged.

\section{Conclusions}
In summary, we have proposed a method to realize DMS with p- and n-type carriers by choosing host semiconductors with a narrow band gap. Using the combined method of DFT and QMC, we describe DMS Mn-doped BaZn$_2$As$_2$,
which has a narrow band gap of 0.2 eV. In addition, we find a nontoxic DMS Mn-doped BaZn$_2$Sb$_2$, whose T$_c$ is expected to be higher than that of Mn-doped BaZn$_2$As$_2$, for which T$_c$ = 230 K, as reported in a recent experiment.
   	
\section*{Acknowledgments}	
The authors acknowledge H. Y. Man, F. L. Ning,  C. Q. Jin, H. Suzuki, and A. Fujimori for many valuable
discussions about the experiments of Mn-doped BaZn$_2$As$_2$. 
   

\end{document}